\documentclass [aps, prb, amssymb, twocolumn, showpacs, superscriptaddress]{revtex4-1}
\usepackage{graphicx,epsfig, amsfonts, amssymb, amsmath}
\usepackage{bm}
\usepackage{times}
\usepackage[colorlinks,citecolor=blue,linkcolor=red]{hyperref}
\usepackage{color}

\begin{document} 

\title{Theory of asymmetric and negative differential magnon tunneling under temperature bias: Towards a spin Seebeck diode and transistor}

\author {Jie Ren}\email{renjie@lanl.gov}
\affiliation{Theoretical Division, Los Alamos National Laboratory, Los Alamos, New Mexico 87545, USA} 
\author{Jian-Xin Zhu}
\affiliation{Theoretical Division, Los Alamos National Laboratory, Los Alamos, New Mexico 87545, USA} 
\affiliation{Center for Integrated Nanotechnologies, Los Alamos National Laboratory, Los Alamos, New Mexico 87545, USA}

\date{\today}

\begin{abstract}
We study the nonequilibrium transport for the asymmetric and negative differential magnon tunneling driven by temperature bias. We demonstrate that  the many-body magnon interaction that makes the magnonic spectrum temperature-dependent is the crucial factor for the emergence of rectification and negative differential spin Seebeck effects in magnon tunneling junctions. When magnonic junctions have temperature-dependent density of states,  reversing the temperature bias is able to give asymmetric spin currents and increasing temperature bias could give an anomalously decreasing magnonic spin current. We show that these properties are relevant for building spin Seebeck diodes and transistors, which could play important roles in controlling information and energy in magnonics and spin caloritronics.
\end{abstract}



\maketitle

\section{Introduction}

In the fast-developing fields of spintronics, \cite{spintronics} magnonics, \cite{magnonics} and spin caloritronics, \cite{Slonczewski, BauerReview} many intriguing effects have been observed. One of the most interesting discoveries is the spin Seebeck effect (SEE), which is a phenomenon that a temperature bias can drive a pure spin current. It has been identified in many kinds of materials, including magnetic metals, \cite{Uchida2008Nature} magnetic semiconductors, \cite{Jaworski2010NatureMat, Breton2011Nature} magnetic insulators, \cite{Uchida2010NatureMat, Uchida2010APL} and nonmagnetic materials with spin-orbit coupling. \cite{Jaworski2012Nature}

The real breakthrough is made by the observation of SSE in magnetic insulators, \cite{Uchida2010NatureMat, Uchida2010APL} which has clearly uncovered that, distinct from spin-dependent Seebeck effect in metallic materials, SSE possesses the unique ability to generate a pure flow of spin angular momentum by mere thermal excitations without moving charge carriers. The thermal-generated pure spin current is carried by excitations of the magnetization, known as magnons, instead of by moving charges.
This is an advantage because charge carriers are often problematic for the thermal design of devices, of which the issue can be avoided by the SSE in insulating magnets without conducting charge currents. It allows us to construct efficient thermoelectric devices upon new principles \cite{Kirihara2012NatureMat} and to realize robust, nondissipative information transmission and energy transfer \cite{Kajiwara2010Nature, TMI} in the absence of Joule heating.
Therefore, the SSE is expected to act as a new method facilitating the functional use of ``waste'' heat and opens a new possibility of spintronics, \cite{spintronics} magnonics, \cite{magnonics} and spin caloritronics,  \cite{Slonczewski, BauerReview} and thus has ignited a new upsurge of research interest \cite{Hinzke2011PRL, Kovalev2012EPL, Jansen2012PRB, Sinova2012NM, Ohnuma2013PRB, Cunha2013PRB, Qu2013PRL, Kikkawa2013PRL} in these fields. 

\begin{figure}
\scalebox{0.5}[0.45]{\includegraphics{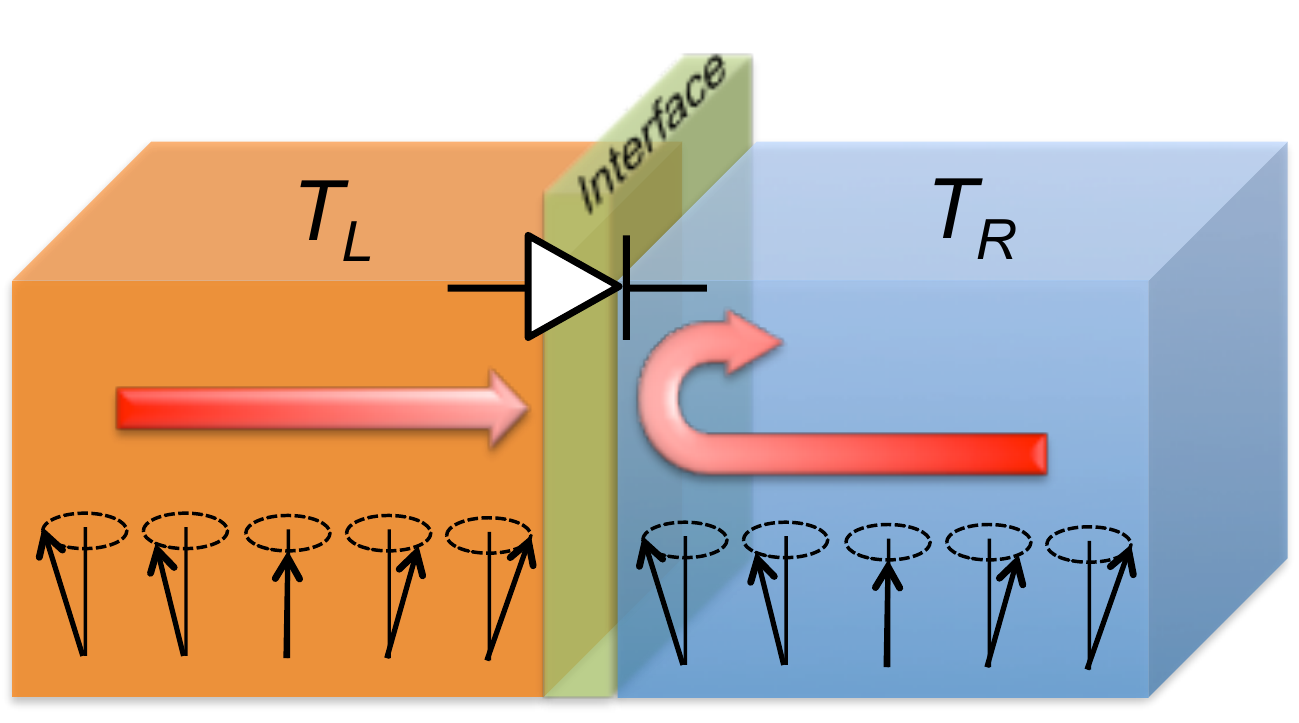}}
\vspace{-3mm}   
\caption{(Color online) Schematic illustration of  the magnon tunneling junction. By acting as a spin Seebeck diode, for $T_L>T_R$ the magnonic spin current is easily flowing from left to right but for $T_R>T_L$ the magnonic spin current from right to left is severely suppressed and even prohibited. 
} \label{fig1}
\end{figure}

A recent work by one of the authors has uncovered the {\it rectification} and {\it negative differential} SSE in a metal-insulating magnetic interface system, \cite{spinSeebeckdiode} where the spin transfer is assisted by the interface electron-magnon inelastic scattering. It was shown that  with the interfacial electron-magnon coupling, reversing the thermal bias is able to give asymmetric spin currents, and increasing thermal bias across the interface abnormally gives an increasing spin current. \cite{spinSeebeckdiode} 
In this work, we study the nonequilibrium transport for the asymmetric and negative differential magnon tunneling driven by temperature bias. We demonstrate in magnon tunneling junctions that the rectification and negative differential SSE emerge as a consequence of  the magnon-magnon interaction that makes the magnonic spectrum {\emph{temperature-dependent}}. This mechanism is different from the one underlying the asymmetric and negative differential SSEs in metal-magnetic insulator interfaces~\cite{spinSeebeckdiode} that needs the {\emph{energy-dependent}} electronic spectrum. We also illustrate the concept of spin Seebeck transistor based on the negative differential SSE. 

The rectification and negative differential electronic transports are fundamental for realizing functional electronic diodes and transistors, which are building blocks of modern electronics. Similarly, the rectification of heat flux and negative differential thermal conductance~\cite{RZ2013PRB2} are also crucial for designing heat diodes and transistors that are as well the fundamental building blocks of phononics. \cite{phononics} Thus, we believe our results are relevant for constructing magnonic and spin caloritronic circuits with efficient spin Seebeck diodes and transistors, which could play crucial roles in controlling energy and information in  magnonics \cite{magnonics} and spin caloritronics. \cite{BauerReview}

We summarize the transport theory of magnon tunneling in a spin tunneling junction in Sec. II. We point out in Sec. IIIA that the many-body interaction is important for rectifying spin Seebeck currents. In Sec. IIIB, we show one example of the spin Seebeck diode with magnon-magnon interaction induced temperature-dependent spectrum. In Sec IIIC, we demonstrate the negative differential SSE in a magnon tunneling junction with temperature-dependent excitation gaps and illustrate the functionality of a spin Seebeck transistor. We conclude in Sec. IV and discuss the possible extension of the present results to phononics of controlling phononic information and thermal energy. \cite{phononics}

\section{Tunneling Hamiltonian and Magnonic Spin Current}

The magnonic spin tunneling junction consists of three parts: When putting together the left and right insulating magnetic segments, the central interface will be formed, which may result from the lattice mismatch, the vacuum, and the depleted- or nonmagnetic regime that acting as the tunneling barrier [see Fig. \ref{fig1}]. Therefore, the total Hamiltonian can be written as:
\begin{eqnarray}
\hat H=\hat H_L+\hat H_I+\hat H_R.
\end{eqnarray}

The left and right insulating magnetic materials, in the low temperature or large spin size limit, are well described by the noninteracting magnon model
\begin{eqnarray}
\hat H_L=\sum_{\bm k\in L}\varepsilon_{\bm k} a^{\dag}_{\bm k}a_{\bm k}, \quad \hat H_R=\sum_{\bm p\in R}\varepsilon_{\bm p} a^{\dag}_{\bm p}a_{\bm p},
\label{eq:freeboson}
\end{eqnarray}
where $a^{\dag}_{\bm k(\bm p)}$ and $a_{\bm k(\bm p)}$ are the creation and annihilation operators of the magnon with momentum $\bm k (\bm p)$ and energy $\varepsilon_{\bm k} (\varepsilon_{\bm p})$ at the left (right) segments. These field operators satisfy the commutation relations for bosons. The left and right magnons are assumed at their own equilibriums with temperature $T_L$ and $T_R$, respectively. Note that since we are interested in the SSE where the pure spin current is induced by the mere temperature bias, we do not consider the possible chemical potential imbalance for the left and right magnons and set them both to zero. In the next section we will discuss more details, where many-body magnon interaction is included.

For the tunneling part, the magnonic spin transfer is described by the tunneling matrix:
\begin{equation}
\hat H_I=\sum_{\bm{kp}}\left(V_{\bm{kp}}a^{\dag}_{\bm k}a_{\bm p}+V^*_{\bm{kp}}a^{\dag}_{\bm p}a_{\bm k} \right),
\end{equation}
which results from the exchange interaction across the interface. Because this term could be small compared to other parts of the total Hamiltonian, it is usually treated by the perturbation theory.

Physically, each magnon carries a unit spin angular momentum $\hbar$. The annihilation of a magnon at one side and the successive creation of a magnon at the other side correspond to the rising of the spin component at one side but lowering the spin component at the other side. In other words, the tunneling magnon current is equivalent to the pure spin down current if the magnetization of the material is defined as up.
The tunneling magnon current operator $\hat I_s$ through the interface is given by $\hat I_s=-\dot{ \hat{N}}_L=\dot{ \hat{N}}_R$, where $\hat N_L=\sum_{\bm k\in L}a^{\dag}_{\bm k}a_{\bm k}$ and $\hat N_R=\sum_{\bm p\in R}a^{\dag}_{\bm p}a_{\bm p}$ are the magnon number operators of the left and right segments, respectively. Using the Heisenberg's equation $\dot{ \hat{N}}_v=\frac{i}{\hbar}[\hat H, \hat N_v]=\frac{i}{\hbar}[\hat H_I, \hat N_v]$, ($v=L,R)$, we can express the tunneling magnon current operator as
\begin{equation}
\hat I_s=\frac{i}{\hbar}\sum_{\bm{kp}}\left(V_{\bm{kp}}a^{\dag}_{\bm k}a_{\bm p}-V^*_{\bm{kp}}a^{\dag}_{\bm p}a_{\bm k} \right).
\end{equation}
The magnonic spin current $I_s$ across the tunneling interface is the average value of the operator $\langle \hat I_s\rangle$, which under the second-order perturbation of the tunneling matrix is calculated through
\begin{equation}
I_s=-\frac{i}{\hbar}\int_{-\infty}^{t}d\tau \left\langle \left[\hat I_s(t), \hat H_I(\tau)\right]\right\rangle,
\end{equation}
where $\hat I_s(t)=e^{i\hat H_0t}\hat I_se^{-i\hat H_0t}$ and $\hat H_I(\tau)=e^{i\hat H_0\tau}H_Ie^{-i\hat H_0\tau}$ are represented in the interaction picture, with $\hat H_0=\hat H_L+\hat H_R$. Substituting the operator $\hat I_s$, we arrive at the single-magnon tunneling current
\begin{equation}
I_s=-\frac{2}{\hbar^2}\text{Im}[G_{\mathrm{tot}}^r(\omega)]_{\omega=0}=\frac{i}{\hbar^2}\left[G_{\mathrm{tot}}^r(\omega)-G_{\mathrm{tot}}^a(\omega)\right]_{\omega=0}, 
\end{equation}
where the retarded (advanced) Green's function reads $G_{\mathrm{tot}}^{r(a)}(\omega)=\int^{\infty}_{-\infty}dt e^{i\omega t}G_{\mathrm{tot}}^{r(a)}(t)$ with $G_{\mathrm{tot}}^{r(a)}(t)=\mp i \Theta(\pm t)\langle [\hat B(t), \hat B^{\dag}(0)] \rangle$ and $\hat B(t)=\sum_{\bm{kp}}V_{\bm{kp}}a^{\dag}_{\bm k}(t)a_{\bm p}(t)$.
Although this tunneling spin current is formulated in terms of the bosonic magnon tunneling and the bosonic Green's functions are defined in a different way from the electronic ones due to the different (anti)commutation relations, we note that it shares the same expression as for the standard result of  fermionic electron tunneling. \cite{Mahanbook} 
The Green's functions are often calculated by evaluating equivalent Matsubara functions of imaginary frequency and successively by the analytic continuation $(i\omega_n\rightarrow \omega \pm i0^+)$.~\cite{Mahanbook} In the following, we alternatively adopt the simple way in the real time and frequency so that the calculations are accessible for the beginner.

To evaluate the tunneling spin current, we use the apparent relation $G_{\mathrm{tot}}^{r}-G_{\mathrm{tot}}^{a}=G_{\mathrm{tot}}^>-G_{\mathrm{tot}}^<$ so that $I_s=-\frac{i}{\hbar^2}\left[G_{\mathrm{tot}}^>(\omega)-G_{\mathrm{tot}}^<(\omega)\right]_{\omega=0}$, where the lesser Green's function in time domain is $G_{\mathrm{tot}}^<(t)=-i\langle \hat B^{\dag}(0) \hat B(t) \rangle$ and the greater Green's function is $G_{\mathrm{tot}}^>(t)=-i\langle  \hat B(t) \hat B^{\dag}(0) \rangle$. In the tunneling process, since the left and right segments of the junction are independent, we then can factorize the Green's function as a product of the corresponding left and right ones. For example, 
\begin{eqnarray}
G_{\mathrm{tot}}^>(t)&=&-i\sum_{\bm{kp}}\sum_{\bm{k'p'}}V_{\bm{kp}}V_{\bm{k'p'}}\left\langle a^{\dag}_{\bm{k}}(t)a_{\bm{p}}(t)a^{\dag}_{\bm{p}}(0)a_{\bm{k}}(0) \right\rangle  \nonumber \\
&=&-i\sum_{\bm{kp}}|V_{\bm{kp}}|^2 \langle a^{\dag}_{\bm{k}}(t)a_{\bm{k}}(0) \rangle \langle a_{\bm{p}}(t)a^{\dag}_{\bm{p}}(0) \rangle  \nonumber \\
&=&i\sum_{\bm{kp}}|V_{\bm{kp}}|^2 G^<_L(\bm{k},-t)G^>_R(\bm{p},t).
\end{eqnarray}
Then, noticing that $G^<_L(\bm{k},t)=-i\langle a^{\dag}_{\bm k}(0) a_{\bm k}(t) \rangle=-iN_L(\varepsilon_{\bm k})e^{-i\varepsilon_{\bm k}t/\hbar}$ and $G^>_R(\bm{p},t)=-i\langle a_{\bm p}(t) a^{\dag}_{\bm p}(0) \rangle= -i [1+N_R(\varepsilon_{\bm p})] e^{-i\varepsilon_{\bm p} t/\hbar}$ where $N_v(\varepsilon)=[\exp({\varepsilon}/{k_BT_v})-1]^{-1}$ is the corresponding Bose-Einstein distribution of the magnon population at the $v$ segment with temperature $T_{v}$, one immediately arrives at
\begin{equation}
G_{\mathrm{tot}}^>(\omega)=-2\pi i\sum_{\bm{kp}} |V_{\bm{kp}}|^2 N_L(\varepsilon_{\bm k})[1+N_R(\varepsilon_{\bm p})]\delta(\omega+\frac{\varepsilon_{\bm k}-\varepsilon_{\bm p}}{\hbar}).
\end{equation}
With the same procedure, one also has
\begin{equation}
G_{\mathrm{tot}}^<(\omega)=-2\pi i\sum_{\bm{kp}} |V_{\bm{kp}}|^2 [1+N_L(\varepsilon_{\bm k})]N_R(\varepsilon_{\bm p})\delta(\omega+\frac{\varepsilon_{\bm k}-\varepsilon_{\bm p}}{\hbar}).
\end{equation}
Therefore, we finally obtain the tunneling spin current:
\begin{equation}
I_s=\frac{2\pi}{\hbar}\sum_{\bm{kp}}|V_{\bm{kp}}|^2\delta(\varepsilon_{\bm k}-\varepsilon_{\bm p}) \left[ N_L(\varepsilon_{\bm k})-N_R(\varepsilon_{\bm p})\right].
\label{eq:Is1}
\end{equation}

For the noninteracting magnon picture within the left and right segments, we have the spectral functions for both sides as $A_L(\bm{k},\varepsilon)=2\pi\delta(\varepsilon-\varepsilon_{\bm k})$ and $A_R(\bm{p},\varepsilon)=2\pi\delta(\varepsilon-\varepsilon_{\bm p})$, so that the magnonic tunneling spin current can be expressed as
\begin{equation}
I_s=\frac{1}{2\pi\hbar} \!\! \int^{\infty}_0 \!\!\!\! d\varepsilon \sum_{\bm{kp}}|V_{\bm{kp}}|^2 A_L(\bm k, \varepsilon)A_R(\bm p, \varepsilon) \left[ N_L(\varepsilon)-N_R(\varepsilon)\right].
\label{eq:Is2}
\end{equation}
Note the density of states (DOS) is related to the spectral functions as $\rho_L(\varepsilon):=\frac{1}{2\pi}\sum_{\bm k}A_L(\bm k, \varepsilon)$ and $\rho_R(\varepsilon):=\frac{1}{2\pi}\sum_{\bm p}A_R(\bm p, \varepsilon)$, when approximating the tunneling matrix element as a constant $|V_{\bm{k}\bm{p}}|\approx|V|$, \cite{Bardeen} one will get
\begin{equation}
I_s=\frac{2\pi}{\hbar} |V|^2 \int^{\infty}_0 \!\!\!\! d\varepsilon \rho_L(\varepsilon)\rho_R(\varepsilon) \left[ N_L(\varepsilon)-N_R(\varepsilon)\right].
\label{eq:Is3}
\end{equation}
This formula shows that the tunneling spin current is determined by the magnon population difference and the overlap between magnon DOS of the left and right segments. In spite of the fact that the spin current is formulated in terms of the bosonic magnon tunneling and the bosonic Green's functions are different from the electronic ones, the expression Eq.  (\ref{eq:Is3}) is still reminiscent of that for the electron tunneling. \cite{Bardeen, Cohen, Schrieffer} The only differences are that here the integral of magnon energy is from $0$ from $\infty$ instead of the energy range $(-\infty,\infty)$ for electrons; the magnon population is characterized by the Bose-Einstein distribution instead of the Fermi-Dirac distribution for electrons. By the substitutions $\sum_{\bm k}\rightarrow\int d\varepsilon_{\bm k}\rho_L(\varepsilon_{\bm k})$ and $\sum_{\bm p}\rightarrow\int d\varepsilon_{\bm p}\rho_R(\varepsilon_{\bm p})$, Eq. (\ref{eq:Is3}) can be also obtained directly from Eq.  (\ref{eq:Is1}).

\section{Results and Discussions}

In the above section, we have summarized the theoretical derivations of the tunneling spin current carried by magnons. It shows clearly that the temperature bias only manifests in the distribution difference of magnons [see Eqs. (\ref{eq:Is1}, \ref{eq:Is2}, \ref{eq:Is3})]. As such, when reversing the thermal bias $T_L \leftrightarrow T_R$, we merely get the reversed spin current without changing magnitude, $I_s\rightarrow -I_s$, i.e., the spin Seebeck diode is absent. In order to obtain the rectification of spin current controlled by the temperature bias, it is necessary for both sides of the tunneling junction to have different responses to temperature change. Therefore, we need to go beyond the noninteracting magnon picture so that the temperature-dependent DOS becomes possible.

\subsection{Many-body interaction effect for rectifying spin Seebeck current}

In many-body systems, the (nonlinear, higher order than quadratic) interaction generally changes the energy spectrum. In terms of the language of Green's functions, the many-magnon interaction manifests as the so-called self energy $\Sigma(\bm k, \varepsilon, T)$ that is generally temperature-dependent. Following the Dyson equation, \cite{NEGFbook} the retarded (advanced) Green's function at temperature $T$ including magnon-magnon (or other many-body) interactions has the form
\begin{equation}
G^{r(a)}=\frac{1}{\varepsilon-\varepsilon_{\bm k}-\text{Re}[\Sigma(\bm k, \varepsilon, T)]\mp i \text{Im}[\Sigma(\bm k, \varepsilon, T)]}.
\label{eq:magnonG}
\end{equation}
Accordingly, the spectral function including the many-magnon interaction obtains the temperature-dependence as well, through the self-energy in Green's functions:
\begin{eqnarray}
&&A(\bm k, \varepsilon, T)=-2\text{Im}[G^r]=i\left[G^r-G^a\right]
\nonumber \\
&&=\frac{-2\text{Im}[\Sigma(\bm k, \varepsilon, T)]}{(\varepsilon-\varepsilon_{\bm k}-\text{Re}[\Sigma(\bm k, \varepsilon, T)])^2+ (\text{Im}[\Sigma(\bm k, \varepsilon, T)])^2}.
\end{eqnarray}
Taking the limit of $\text{Im}\Sigma \rightarrow 0$ as in the mean-field approach, we retrieve the delta-function-type spectral function $A(\bm k, \varepsilon, T)=2\pi\delta(\varepsilon- \tilde{\varepsilon}_{\bm k, T})$ as for the noninteracting case. The temperature-dependence now enters in the renormalized effective free quasiparticle energy $\tilde{\varepsilon}_{\bm k, T}={\varepsilon}_{\bm k}-\text{Re}[\Sigma(\bm k, \varepsilon, T)]$. Thereupon, the magnon 
DOS becomes also temperature-dependent $\rho(\varepsilon, T):=\frac{1}{2\pi}\sum_{\bm k}A_L(\bm k, \varepsilon, T)=\sum_{\bm k}\delta(\varepsilon-\tilde{\varepsilon}_{\bm k, T})$. As such,  Eq. (\ref{eq:Is3}) is modified by the many-body interaction, as
\begin{equation}
I_s=\frac{2\pi}{\hbar} |V|^2 \int^{\infty}_0 \!\!\!\! d\varepsilon \rho_L(\varepsilon, T_L)\rho_R(\varepsilon, T_R) \left[ N_L(\varepsilon)-N_R(\varepsilon)\right].
\label{eq:Is4}
\end{equation}

Generally, the left and right magnetic materials of the tunneling junction can be different, i.e., their magnon DOS have different responses for temperature change. After reversing the thermal bias, i.e., exchanging the temperatures $T_L\leftrightarrow T_R$, we generally have the different DOS-overlaps (or, say, different spectral-overlaps) $\rho_L(\varepsilon, T_R)\rho_R(\varepsilon, T_L)\neq \rho_L(\varepsilon, T_L)\rho_R(\varepsilon, T_R)$. As a result, the reversed spin current will have different magnitudes under reversing thermal bias. If the density-overlap $\rho_L(\varepsilon, T_R)\rho_R(\varepsilon, T_L)$ is much larger or smaller than $\rho_L(\varepsilon, T_L)\rho_R(\varepsilon, T_R)$, we will obtain a spin Seebeck diode. That is, in one direction the temperature bias could produce a considerable spin current but in the opposite direction the temperature bias produces less spin current inefficiently, as illustrated in Fig.~\ref{fig1}. 

Even if the left and right magnetic materials are identical, in some cases the density-overlap $\rho_L(\varepsilon, T_L)\rho_R(\varepsilon, T_R)$ may decrease as increasing the thermal bias $|T_L-T_R|$. When the decreasing of the density-overlap surpasses the increasing of $[N_L-N_R]$, we will have the negative differential SSE. That is, the magnonic spin current decreases with an increasing thermal bias.

Therefore, the many-body interaction induced temperature-dependence of the DOS (as well as the temperature-dependent spectral functions) is the key to achieve the rectification and negative differential transport of spin current by controlling temperature bias.

\subsection{Rectifying spin Seebeck current with magnon-magnon interaction}

In the following, let us use a simple standard model to illustrate how the magnon-magnon interaction induces the temperature-dependent magnon energy   \cite{Oguchi, Bloch} that in turn leads to the possible spin Seebeck diode. Without loss of generality, we focus on the ferromagnets. The results and discussions are readily generalized to the antiferromagnets. \cite{Oguchi}
 
We first look into the left insulating magnetic material, which is conventionally described by a Heisenberg lattice: 
 \begin{eqnarray}
\hat H_L&=&-J_L\sum_{\langle i,j \rangle}{\bm S}_i\cdot{\bm S}_j \nonumber \\
&=&-J_L\sum_{\langle i,j \rangle}[{\frac{1}{2}S^+_iS^-_j+\frac{1}{2}S^+_iS^-_j}+S^z_iS^z_j],
 \end{eqnarray}
where $S^{\pm}_j=S^x_i\pm S^y_i$ is the raising (lowering) operator for the spin at site $j$ and ${\langle i,j \rangle}$ denotes the nearest-neighbor bond with exchange coupling strength $J_L$. This Heisenberg exchange Hamiltonian can be expanded in increasing powers of the magnon operators by the Holstein-Primakoff transformation, \cite{HP}  which maps the  spin operators into bosonic magnons through $S^{+}_j=\sqrt{2S_L-a^{\dag}_ja_j}\,a_j$, $S^{-}_j=a^{\dag}_j\sqrt{2S_L-a^{\dag}_ja_j}$, $S^z_j=S_L-a^{\dag}_ja_j$, where $S_L$ is the spin size of the left material. 
 Clearly, the creation (annihilation) of a magnon $a^{\dag}_j(a_j)$ at site $j$ corresponds to  that spin points less (more) in $z$ component so that each magnon carries an spin angular momentum of $-1$ (associated with a magnetic moment). Thus, the tunneling magnons driven by temperature bias are responsible for the Seebeck spin current.

When keeping up to the lowest order of two-body magnon-magnon interaction, we have the truncated expansion:
 \begin{eqnarray}
  \hat H_L &=& J_LS_L \sum_{\langle ij \rangle}(a^{\dag}_ia_i +a^{\dag}_ja_j -a^{\dag}_ia_j - a^{\dag}_ja_i )  \nonumber \\ 
&+& \frac{J_L}{4}(a^{\dag}_ia^{\dag}_ia_ia_j + a^{\dag}_ia^{\dag}_ja_ja_j + a^{\dag}_ia^{\dag}_ja_ia_i    \nonumber\\
&+& a^{\dag}_ja^{\dag}_ja_ia_j - 4 a^{\dag}_ia_ia^{\dag}_ja_j ) +  \mathcal O\left( \frac{1}{S_L} \right),
 \end{eqnarray}
where we dropped an irrelevant constant term corresponding to the ground-state energy.
Since the Holstein-Primakoff transformation requires the constraint ${\langle a^{\dag}_ja_j\rangle}/{(2S_L)}< 1$, the above expansion is justified, 
as in the large spin limit or low temperatures. After performing the Fourier transform into the momentum space, one gets
\begin{eqnarray}
\hat H_L&=&\sum_{\bm k}\varepsilon_{\bm k}a^{\dag}_{\bm k}a_{\bm k}+\frac{J_L}{4n}\sum_{\bm k \bm k' \bm k''}\sum_{\bm{\delta r}} a^{\dag}_{\bm k}a^{\dag}_{\bm k'}a_{\bm k''}a_{\bm k+\bm k'-\bm k''}  \nonumber \\
&\times&(e^{-i{\bm k}\cdot{\bm{\delta r}}} \! + \! e^{-i({\bm k+\bm k'-\bm k''})\cdot\bm {\delta r}} \! - \! 2e^{-i{(\bm k'-\bm k'')}\cdot{\bm{\delta r}}}), 
\end{eqnarray}
where $n$ is the total number of sites, $\bm{\delta r}$ denotes the nearest-neighbor vector, and the energy spectrum has the dispersion
$\varepsilon_{\bm k}=J_LS_L\sum_{\bm {\delta r}} (1-e^{-i{\bm k}\cdot{\bm{\delta r}}})$. To handle the two-body magnon interaction, we adopt the standard mean-field approximation and keep the Hartree-Fock terms. Finally, after some algebra, one arrives at the quasifree magnon Hamiltonian, as in Ref. \onlinecite{Bloch}:
\begin{equation}
\hat H_L=\sum_{\bm k} \alpha_L(T_L)  \varepsilon_{\bm k} a^{\dag}_{\bm k}a_{\bm k},
\end{equation}
where all the magnon-magnon interactions are renormalized into the temperature-dependent factor 
\begin{equation}
\alpha_L(T_L)=1-\frac{1}{2J_LS^2_Lnz}\sum_{\bm k}\frac{\varepsilon_{\bm k}}{\exp \left(\frac{\alpha(T_L)\varepsilon_{\bm k}}{k_BT_L}\right)-1},
\label{eq:aa}
\end{equation}
with $z$ the lattice coordinate number.
It requires self-consistency to solve $\alpha_L(T_L)$ and the analytic expression is absent. However, to the lowest order we can set the factor at the right-hand side as $1$ and use the long-wave limit $\varepsilon_{\bm k}\approx J_LS_La^2{\bm k}^2$ for the three-dimensional (3D) cube lattice with $a$ the lattice constant, then one can integrate Eq.~(\ref{eq:aa}) and readily obtain
\begin{equation}
\alpha_L(T_L)=1-\frac{\zeta({5}/{2})}{64 S_L  \pi^{3/2}}\left( \frac{k_BT_L}{J_LS_L} \right)^{5/2},
\label{eq:aa1}
\end{equation}
with $\zeta(\cdot)$ denoting the Zeta function.
Accordingly, the temperature-dependent magnon DOS for the 3D cube lattice is obtained as
\begin{equation}
\rho_L(\varepsilon, T_L)=\frac{1}{(2\pi)^2}\sqrt{\frac{\varepsilon}{[J_LS_L\alpha_L(T_L)]^3}} \; ,
\label{eq:DOS}
\end{equation}
with the total number of spins being normalized. 
Same expressions also apply for the right insulating magnetic materials with the subscript interchange $L\rightarrow R$. 

\begin{figure}
\scalebox{0.4}[0.4]{\includegraphics{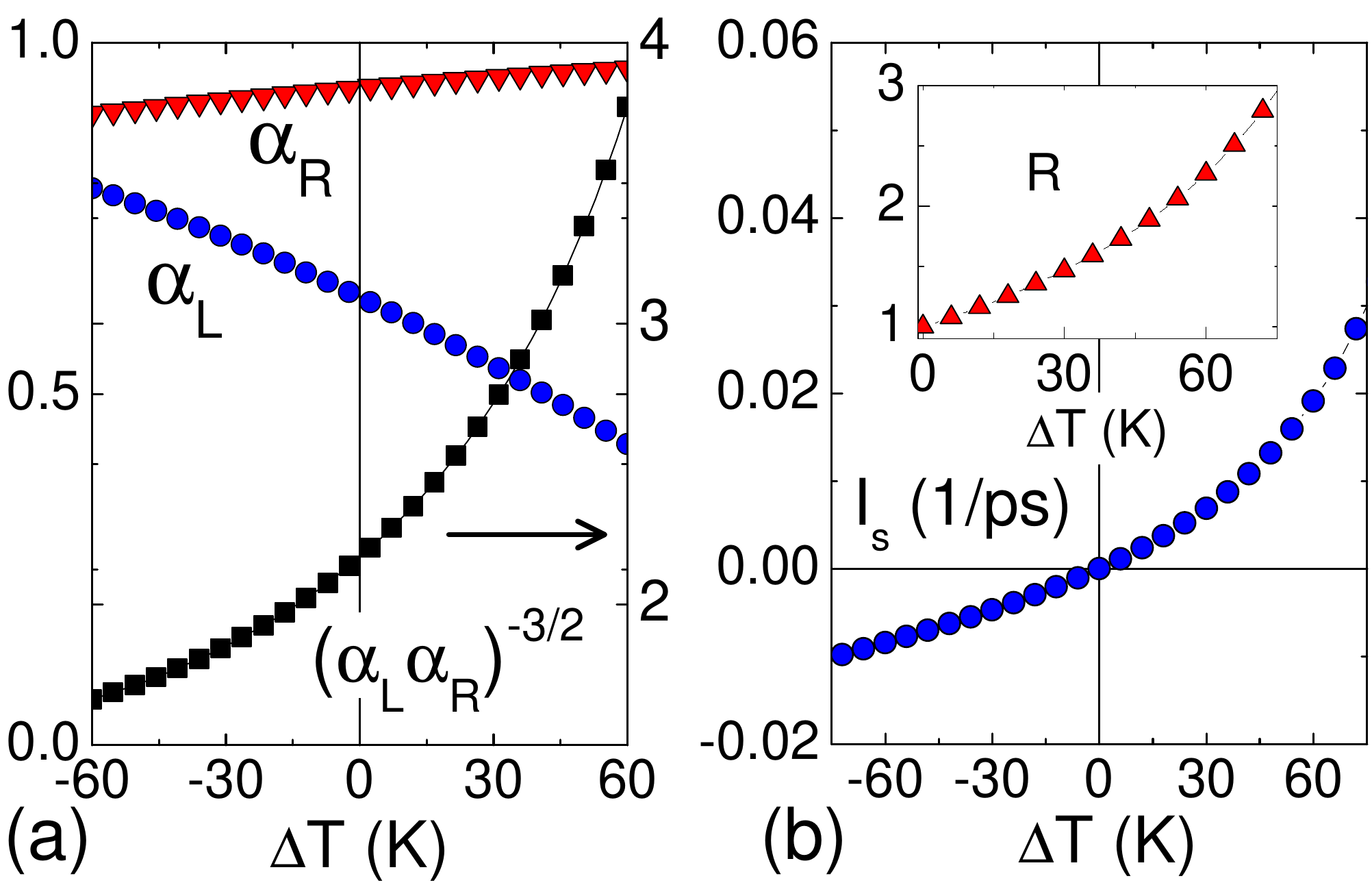}}
\vspace{-4mm}   
\caption{(Color online) (a) {\bf Different temperature responses of the left and right magnonic segments}. The DOS-overlap $\rho_L(\varepsilon, T_L)\rho_R(\varepsilon, T_R)\sim[\alpha_L(T_L)\alpha_R(T_R)]^{-3/2}$ shows the asymmetric behavior. (b) {\bf Asymmetric thermal-generated spin current in a spin Seebeck diode.} Inset shows the rectification ratio defined as $R:=|I_s(\Delta T)/I_s(-\Delta T)|$. Material parameter values are $J_L=0.1$ meV, $J_R=0.2$ meV, $S_L=S_R=14$, which are comparable with the estimated ones from typical magnetic insulators, cf. YIG (Y$_3$Fe$_5$O$_{12}$) in Refs. \onlinecite{YIG1, YIG2, YIG3}. Other parameters are $V=0.1$ meV, $T_L=T_0+\Delta T$, $T_R=T_0-\Delta T$ with $T_0=300$ K. To meet the constraint ${\langle a^{\dag}_{\bm k}a_{\bm k}\rangle}/{(2S_{L,R})}< 1$, we set a magnon gap of $5$ meV, which however will not affect the data plotted here.
} \label{fig2}
\end{figure}

From Eq.~(\ref{eq:aa1}), we know that increasing temperature $T_v$ will decrease $\alpha_v$ that in turn increases the DOS $\rho_v$ through Eq.~(\ref{eq:DOS}) [see also Fig.~\ref{fig2}(a)]. This is understandable because when temperature increases, more magnons are excited so that their interactions effectively soften the exchange stiffness through the decreasing factor $\alpha_v$, which equivalently increases the magnon population. As a consequence, the DOS $\rho_v$ increases (decreases) as temperature $T_v$ increases (decreases) and the increasing (decreasing) rate is a function of $J_v$ and $S_v$.
Therefore, when the left and right segments are made of different magnetic materials, i.e., $J_L \neq J_R$, $S_L \neq S_R$, or different spin lattice structures, their DOS then generally possess different responses to the temperature change through Eq. (\ref{eq:DOS}), which makes the spin Seebeck diode possible.

Figure \ref{fig2}(b) illustrates one example of the asymmetric SSE: in the positive thermal bias $\Delta{T}>0$ ($T_{L,R}=T_0\pm\Delta{T})$, the thermal-generated spin current is considerable while it is suppressed in the negative thermal bias $\Delta{T}<0$. This is reasonable by looking into the DOS-overlap $\rho_L(\varepsilon, T_L)\rho_R(\varepsilon, T_R)\sim[\alpha_L(T_L)\alpha_R(T_R)]^{-3/2}$, which as shown in Fig.~\ref{fig2}(a) is asymmetric for positive and negative thermal bias as the consequence of different responses of $\alpha_L$ and $\alpha_R$ to the temperature change. The behavior of the DOS-overlap is consistent with that of the Seebeck spin current, displayed in Fig.~\ref{fig2}(b).
Naturally, the rectification ratio $R:=|I_s(\Delta T)/I_s(-\Delta T)|$ increases as enlarging the thermal bias [see the inset of Fig.~\ref{fig2}(b)]. Although the rectification ratio $R$ shows that it is not as good as a perfect diode at small thermal bias, the example demonstrates in principle the feasibility of the spin Seebeck diode. 

Note here the rectification of SSE emerges from the {\emph{temperature-dependent}} magnonic spectrum due to the magnon-magnon interaction. It applies for the pure magnonic tunneling system. This mechanism is different from the asymmetric SSE in hybrid metal-magnetic insulator systems,~\cite{spinSeebeckdiode} where the interfacial magnon-electron scattering and the {\emph{energy-dependent}} electronic spectral density are crucial. Although we exemplify the spin Seebeck rectification with only the lowest-order magnon-magnon interaction, at high temperatures more magnons will be excited and high-order magnon interactions will be manifested. As such, the temperature-dependence of the magnon DOS will become stronger, which then magnifies the response difference of the magnon DOS of the two segments to temperature changes. Therefore, more and stronger (high-order) magnon interactions are preferable for realizing efficient spin Seebeck diode.

\subsection{Negative differential spin Seebeck effect and spin Seebeck transistor}

In the above example, we have illustrated that the possible spin Seebeck diode effect  can result from the temperature-dependent spectra induced by the magnon-magnon interaction with however temperature-independent exchange couplings. In fact,  in real situations, the exchange interaction itself $J(T)$ may be sensitive to temperature changes, which in turn also has the contribution to the spin Seebeck diode effect. Beside the exchange coupling, the dipolar interaction will have a similar effect. In particular, the dipolar interaction as well as the anisotropic exchange coupling usually opens a gap for the magnon spectrum, which thus makes the magnon gap temperature-dependent; see Refs. \onlinecite{exp1, exp2, exp3} for experimental examples. The temperature-dependent magnon gap not only contributes to the spin Seebeck diode effect but also makes the negative differential SSE possible; that is,  increasing (decreasing) the thermal bias gives a decreasing (increasing) spin current. 

As an example, we consider the two materials at both sides of the tunneling interface are both 3D cube lattices with finite magnon gaps so that their DOS can be obtained as
\begin{equation}
\rho_v(\varepsilon, T_v)=\frac{1}{\pi}\int^{\infty}_0 \!\!\! d x [\mathcal J_0(2J_vS_vx)]^3\cos[x(\varepsilon-\Delta_v-6J_vS_v)],
\label{eq:DOS3D}
\end{equation}
where $\mathcal J_0(\cdot)$ denotes the Bessel function of the first kind and $\Delta_v$ ($v=L, R$) is the magnon gap of the $v$ side, which we assume linearly increases with temperature $\Delta_v=\gamma_v T_v$, as indicated by the experiments in Refs. \onlinecite{exp1, exp2, exp3}. We would like to iterate that although here we use a phenomenological temperature-dependence of magnon gap, the microscopic reason relies on the general mechanism of the many-body magnon interaction (or other particle-magnon interactions) induced temperature-dependence. We focus on the model study at present, which in principle is sufficient to demonstrate the possible nontrivial properties. Future works should be continued from the aspect of first-principle or {\it ab initio} theory for more realistic material calculations.

\begin{figure}
\scalebox{0.4}[0.4]{\includegraphics{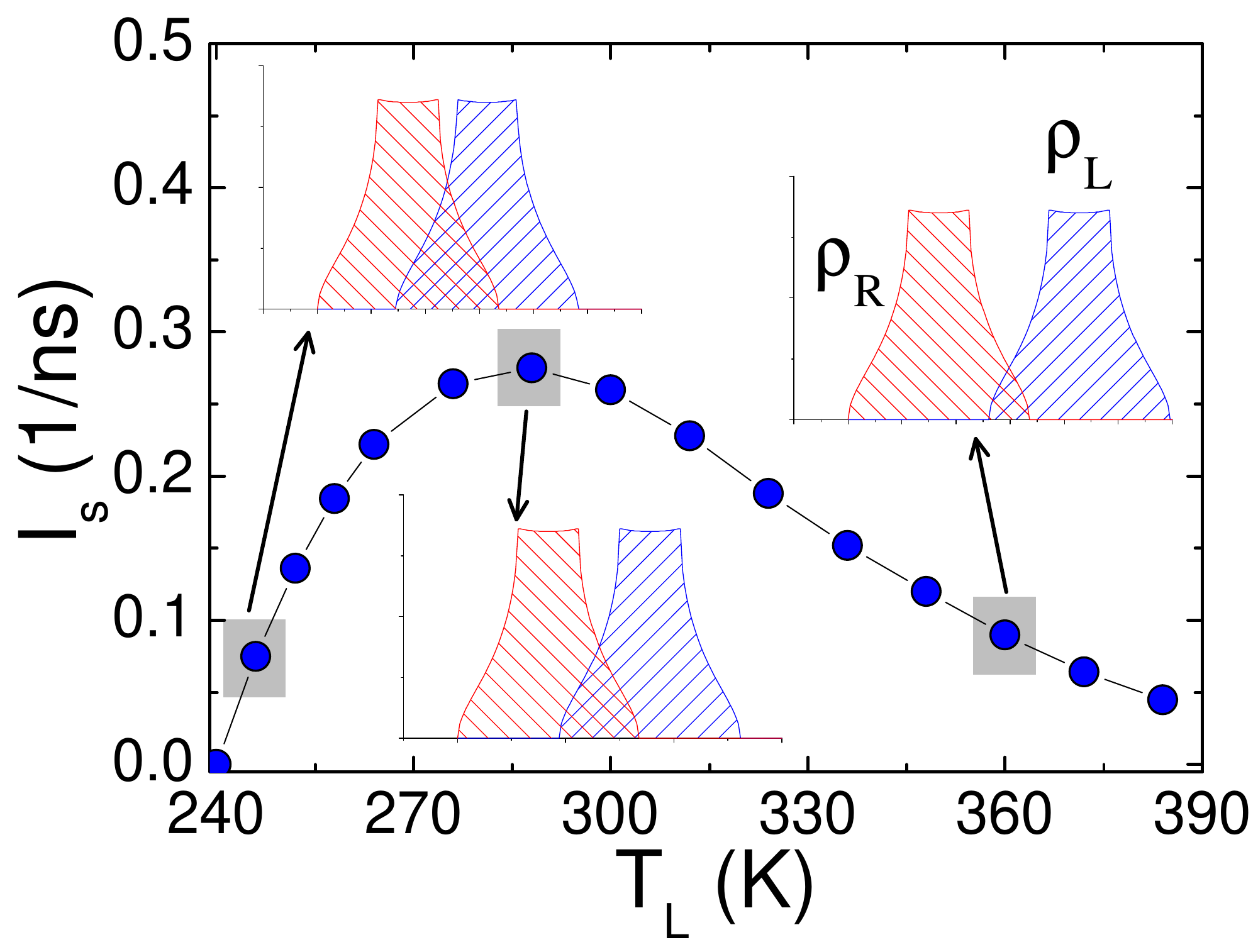}}
\vspace{-4mm}   
\caption{(Color online) {\bf Negative differential spin Seebeck effect.} The insets show that the DOS-overlap decreases with increasing the thermal bias, which causes the behavior of negative differential conductance. Material parameter values are $J_L=J_R=0.1$ meV, $S_L=S_R=14$ for 3D cube lattices that are comparable with the estimated ones from typical magnetic insulators, cf. YIG in Refs. \onlinecite{YIG1, YIG2, YIG3}, and $V=0.1$ meV. The right segment has a fixed magnon gap $\Delta_R=5$ meV with a fixed temperature $T_R=240$ K. The magnon gap for the left material is assumed as $\Delta_L=\gamma_L T_L$ with $\gamma_L=0.05$ meV/K. The three DOS profiles are calculated from Eq.~(\ref{eq:DOS3D}).} 
\label{fig3}
\end{figure}

The corresponding negative differential SSE is illustrated in Fig. \ref{fig3}.
When $T_L$ increases larger than $T_R$, the spin Seebeck current first increases as expected but then counterintuitively  decreases with further increasing thermal bias, so called \emph{negative differential SSE}.  
This anomalous behavior can be understood with the insets of the figure, where the corresponding magnon DOS of the left and right materials are depicted through calculating Eq.~(\ref{eq:DOS3D}):
With increasing the thermal bias, the DOS of two sides are drawn away from each other, which in turn decreases the DOS-overlap in Eq. (\ref{eq:Is4}); When the DOS-overlap decreasing surpasses the increasing of thermal bias, the spin current starts to decrease although with increasing the bias, and the negative differential SSE then emerges. This mechanism for the negative differential SSE is different from the one discussed in Ref. \onlinecite{spinSeebeckdiode}, where both the interfacial electron-magnon coupling and the strong energy-dependent electronic DOS play the crucial role.

\begin{figure}[b]
\scalebox{0.5}[0.5]{\includegraphics{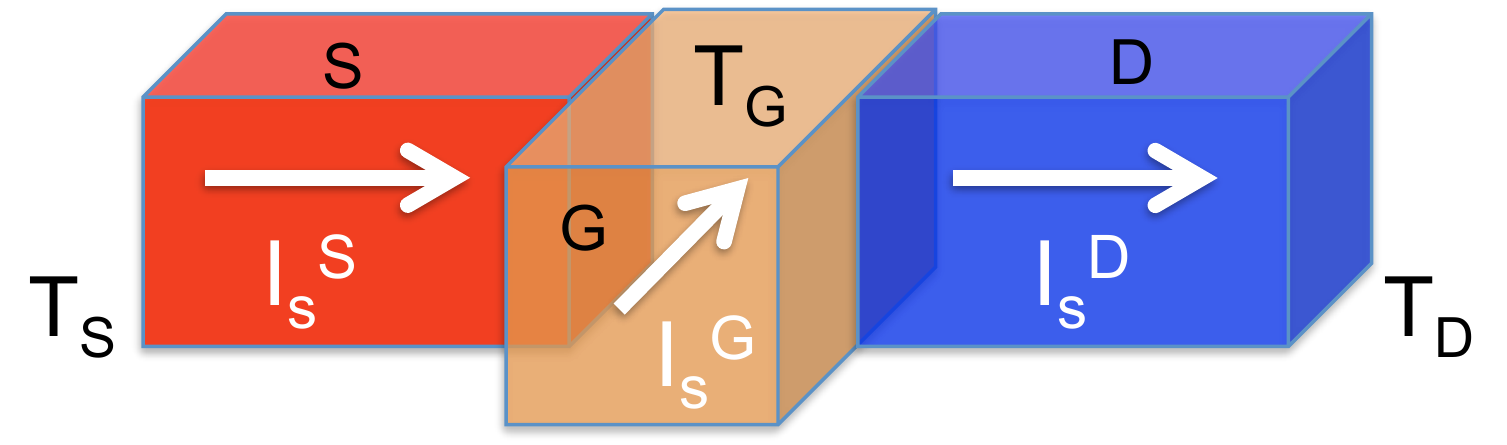}}
\vspace{-5mm}   
\caption{(Color online) Schematic illustration of the concept of a spin Seebeck transistor.} 
\label{fig4}
\end{figure}

Now, based on the negative differential SSE, let us illustrate the concept of spin Seebeck transistor following the spirit of phononics \cite{phononics}:
Similar to the electronic counterpart of a field-effect transistor (FET), the spin Seebeck transistor is composed of three parts: The source (S), the gate (G), and the drain (D), as depicted in Fig. \ref{fig4}, which may then connect to external spin or magnon circuits. The temperatures at the source and drain are fixed with $T_S>T_D$ so that the Seebeck-effect-generated magnonic spin current flows from the source to the drain. A third temperature $T_G$ at the gate side is tunable to control the spin currents at the source, gate, and drain, with $I_s^{S}=I_s^{D}+I_s^{G}$. The amplification ability of a spin Seebeck transistor can be characterized as the change rate of the gain spin current at the drain upon the change of the control current at the gate, which is expressed as 
\begin{equation}
\beta_{\text{Amp}}=\left| \frac{\partial I_s^D}{\partial I_s^G} \right|=\left| \frac{\partial I_s^D}{\partial (I_s^D-I_s^S)} \right|=\left| \frac{\mathcal G_D}{\mathcal G_D+\mathcal G_S} \right|,
\end{equation}
where $\mathcal G_S=-\partial I_s^S/\partial T_G$ and $\mathcal G_D=\partial I_s^D/\partial T_G$ are the differential spin Seebeck conductance for the source and drain segments, respectively. It is straightforward to see that only when either negative differential spin Seebeck conductance $\mathcal G_S<0$ or $\mathcal G_D<0$ is present then $\beta_{\text{Amp}}$ emerges to be larger than unity. Therefore, the negative differential SSE is crucial for realizing  a spin Seebeck transistor to amplify a spin current with a weak input signal, as the FET in modern electronics and the thermal transistor in phononics. \cite{phononics}

\section{Conclusions}

In summary, we have studied the nonequilibrium transport for the asymmetric and negative differential magnon tunneling driven by temperature bias. We have demonstrated that in magnon tunneling junctions, the rectification and negative differential spin Seebeck effects are emerging as a consequence of  the many-body magnon interaction that makes the magnonic spectrum temperature-dependent, which are then used to build spin Seebeck diodes and transistors. Considering the fact that the magnon carries not only the spin angular momentum but also the energy [see, e.g., Ref. \onlinecite{TMI}], the present results further indicate the potential of a magnon tunneling junction acting as a thermal diode and/or a thermal transistor. 

Although in calculations we normalized the number of spin at the interface, the size effect will not influence the asymmetric and negative differential magnon tunneling behaviors, but only affects the magnitude of the spin current. We can restore the interface size by denoting the number of spin at the tunneling interface as $N_s$. In this way, $\rho_v\rightarrow{N}_s\rho_v$ and $V\rightarrow{V}/\sqrt{N_s}$. Thus, the tunneling spin current will be magnified as $I_s\rightarrow{N_s}I_s$. Therefore, in macroscopic magnon tunneling junctions, the spin current can be sufficiently large and much easier for measurements and applications. 

Extension of the asymmetric and negative differential tunneling to phononics for the phonon diode and transistor \cite{phononics} is also possible, but should be done with care. Although the phonon and magnon are both bosons, phonons have different physics in the real space compared to magnons.
For phonons, the left and right side can still be described by the free bosonic gas model as in Eq. (\ref{eq:freeboson}), but then the tunneling matrix is modified to
\begin{equation}
\hat H_I=\sum_{\bm k \bm p} V_{\bm k \bm p} (a_{\bm k}+a^{\dag}_{-\bm k})(a_{\bm p}+a^{\dag}_{-\bm p}),
\end{equation}
because the phonon tunneling is caused by the bilinear coupling of atomic displacements in the real space. Also, the retarded (advanced) Green's function for phonons, instead of Eq. (\ref{eq:magnonG}), has a different form \cite{Mahanbook}:
\begin{equation}
G^{r(a)}=\frac{2 \omega_{\bm k}}{\omega^2-\omega_{\bm k}^2-2\omega_{\bm k}\text{Re}[\Sigma(\bm k, \omega, T)]\mp i 2\omega_{\bm k}\text{Im}[\Sigma(\bm k, \omega, T)]},
\end{equation}
which is also due to the fact that the physics of the phononic Green's functions in real space are expressed in terms of the displacement-displacement correlation. But the underlying mechanism should not change, that is, the many-body interaction induces temperature-dependent self-energies, which in turns makes the quasiparticle spectrum temperature-dependent. As a consequence, the rectification and negative differential thermal conductance will be emergent and the phonon diode and transistor become possible. 

Recent studies also imply that the phonon-drag plays an important role in SSE. \cite{Xiao2010PRB, Adachi2010APL, Jaworski2011PRL, Uchida2011NatureMater, Jaworski2012Nature, Agrawal, Schreier} Thus taking account of the effect of nonequilibrium phonons on asymmetric and negative differential SSE will be an interesting future topic. We therefore believe that by integrating the phononics \cite{phononics} with spintronics, \cite{spintronics} magnonics, \cite{magnonics} and spin caloritronics, \cite{BauerReview} there are more opportunities to achieve the smart control of energy and information in low-dimensional nanodevices.

\begin{acknowledgments}
{This work was supported by the National Nuclear Security Administration of the US DOE at LANL under Contract No. DE-AC52-06NA25396, and the LDRD Program at LANL (J.R.) and in part by the Center for Integrated Nanotechnologies, a U.S. DOE Office of Basic Energy Sciences user facility (J.-X.Z.).}
\end{acknowledgments}


\end{document}